\documentclass[11pt]{article}
\usepackage{graphicx}
\usepackage[font={small,it}]{caption} 
\usepackage{amsopn}

\DeclareMathOperator{\ccg}{ccg}
\title{\textbf{Clique Topology Reveals Intrinsic Geometric Structure in Neural Correlations: An Overview}}
\author{David Cox}
\date{May 2016}
\begin{document}

\maketitle

\section*{Introduction}
Gusti et al. present clique topology -- a novel matrix analysis technique to extract structural features from neural activity data that contains hidden nonlinearities. This represents a significant improvement upon the previous eigenvalue-based approaches that fail under similar conditions. Structural features can yield significant insight into neurological processes.\\

Their findings are verified by showing that the geometric structure of neural correlations in the rat hippocampus is determined by hippocampal circuits rather than being a consequence of positional coding.

\section*{Background}

\subsection*{Clique topology and Betti curves}

Clique topology is a central theme in this paper and is best explained by breaking it down into two terms. A clique is a complete graph, $C$ that is a subgraph of a directed graph $G = (V,E)$. Alternatively, a clique in an undirected graph $G$ is a subset of vertices, $C \subseteq V$ where every pair of distinct vertices are adjacent. Topology in this case refers to the study of the orientation of clique connections in an order complex.\\

\pagebreak

\vspace*{\fill}

For a matrix, $A$, an \textit{order complex} is defined as a subsequence of graphs, ${G_0 \subseteq G_1 \subseteq ... \subseteq G_p}$ such that $G_0$ is the graph having $N$ vertices and no edges, $G_1$ has a single edge $i,j$ corresponding to the highest off-diagonal matrix value $A_{ij}$, and with each subsequent graph having an additional edge for the next-highest off-diagonal matrix entry for any $N \times N$ symmetric matrix $A$ with constant and distinct order of entries. The following figure from the paper illustrates this concept. \\

\begin{figure}[h!]
\centerline{\includegraphics[scale=0.4]{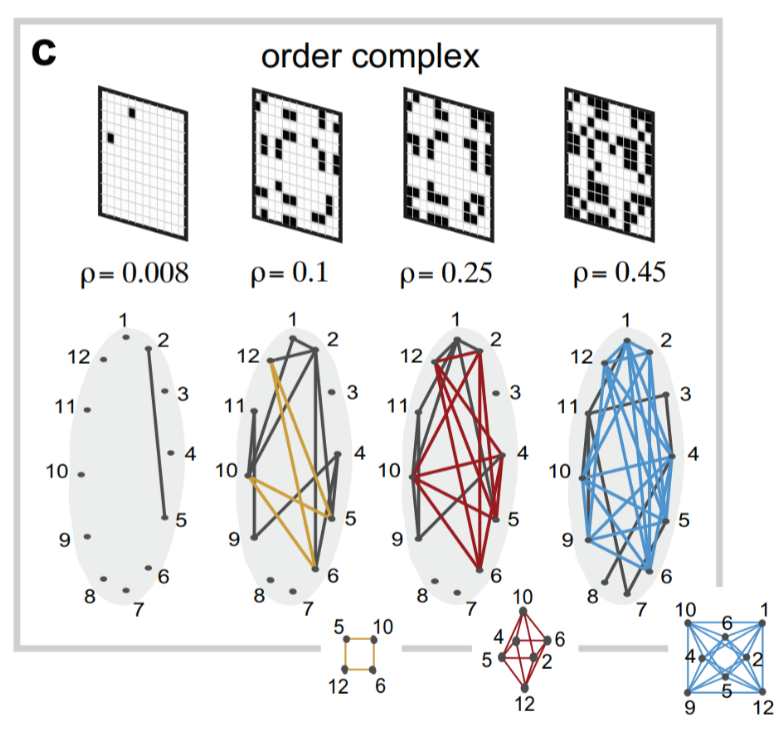}}
  \caption{(Top) The order complex of A is represented as a sequence of binary adjacency matrices, indexed by the density  of non-zero entries. (Bottom) Graphs corresponding to the adjacency matrices. Minimal examples of a 1-cycle (yellow square), a 2-cycle (red octahedron) and a 3-cycle (blue orthoplex) appear at ${\rho = 0.1}$, 0.25, and 0.45, respectively. }
\end{figure}
\vfill

\pagebreak
\enlargethispage{2cm}
As connections are introduced to the order complex, non-contractible holes, or cycles, appear. Cycles are created, modified, and eventually destroyed as the edge density of cliques increases. Betti numbers are used to quantify this behavior. A Betti number $\beta_m$ is a function of the number of cycles in a graph $G$ and the edge density $\rho$. Considering the complete set of edge densities for all $m$-cycles allows for the generation of Betti curves, which provide a summary of the topological features of a given matrix.\\

To compare Betti curves to a control matrix, Gusti et al. introduce the concept of integrated Betti values, $\overline{\beta}_m$.\\

\[\overline{\beta}_m = \int_0^1 \beta_m (\rho) d \rho\]

The values $\beta_1$, $\beta_2$, and $\beta_3$ were computed for each data set. An example for random matrices is shown in the following figure from the original publication.\\

\begin{figure}[h!]
\centerline{\includegraphics[scale=0.35]{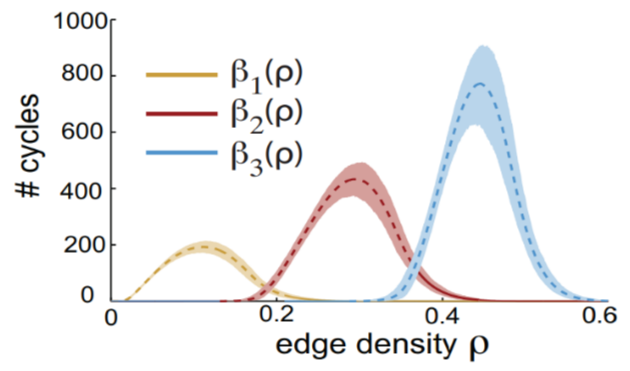}}
  \caption{For a distribution of 1000 random $N \times N$ symmetric matrices ${(N = 88)}$, average Betti curves $\beta_1(\rho),\;  \beta_2(\rho),$ and $\beta_3(\rho)$ are shown (yellow, red, and blue dashed curves), together with 95\% confidence intervals (shaded areas).
}
\end{figure}

Matrices containing random data have Betti curves distinct from those containing geometric data. Random data produces unimodal Betti curves that increase as a function of edge density ($\rho$), while geometric data results in Betti curve values that decrease as a function of $\rho$ as they contain significantly fewer cycles. Betti curves for distributions of geometric matrices are shown in the figure below.\\ 

Note the significant difference in peak values from the previous figure.\\
\vfill

\pagebreak

\vfill
\enlargethispage{2cm}

\begin{figure}[h!]
\centerline{\includegraphics[scale=0.27]{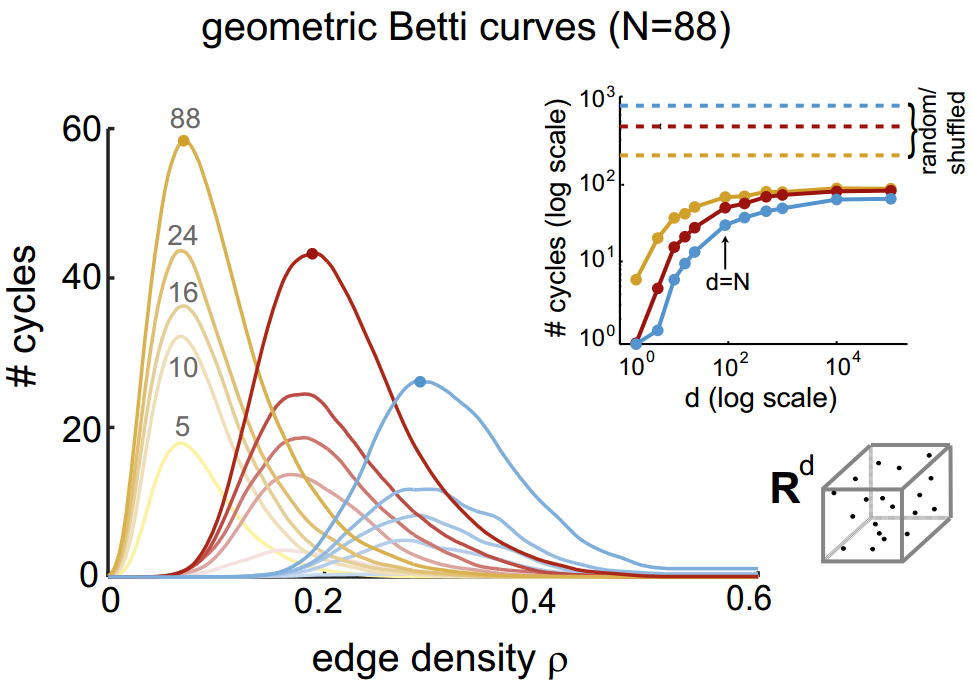}}
  \caption{Betti curves for distributions of geometric matrices ${(N = 88)}$ in dimensions $d$ = 5, 10, 16, 24, and 88. Mean Betti curves $\beta_1(\rho), \; \beta_2(\rho),$ and $\beta_3(\rho)$ are shown (yellow, red, and blue curves), with darker (and higher) curves corresponding to larger $d$.}
\end{figure}

\subsection*{Computation of pairwise correlation matrices}

Neural connectivity data is often represented as a matrix whose entries $C_{ij}$ correspond to the strength of connection between pairs of neurons, with strong connections indicating a high degree of \textit{neural correlation}. The clique topology technique serves to extract structural features from this data.\\

In the following figure, neural connectivity data is represented as ``spike trains'' -- time-series electrical signals recorded from individual neurons. 

\begin{figure}[h!]
\centerline{\includegraphics[scale=0.25]{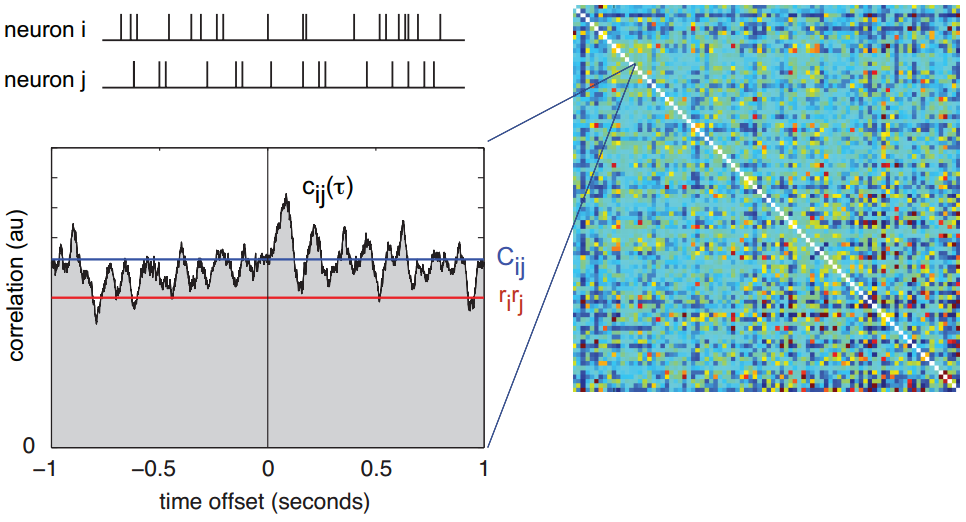}}
  \caption{Computation of pairwise correlation matrices from spike train data.}
\end{figure}
\vfill

\pagebreak

\vspace*{\fill}
For a pair of spike trains $\left\{ t_\ell^i \right\}_{\ell=1...n_j}$ for neurons $i$ and $j$ (top of figure), the cross-correlogram $\ccg_{ij}(\tau)$ is computed as 

\[ \ccg_{ij}(\tau) = \frac{1}{T} \int_0^T f_i(t)f_j(t+\tau)dt, \]

where $f_i(t)=\sum_{\ell=1}^{n_i}\delta \left( t - t_\ell^i \right)$ is the instantaneous firing rate of the $i$-th neuron. The graph in Figure 4 shows a smoothed $\ccg_{ij}(\tau)$ as a black curve along with the expected value of the cross-correlogram, $r_i r_j$ in red for uncorrelated spike trains with matching firing rates, $r_i = \frac{n_i}{T}$. The blue line illustrates the pairwise correlations $C_{ij}$ with timescale $\tau_{\max}$, which were computed as 

\[ C_{ij} = \frac{1}{\tau_{\max} r_i r_j} \max \left( \int_0^{\tau_{\max}} \ccg_{ij} (\tau) d \tau, \int_0^{\tau_{\max}} \ccg_{ji} (\tau) d \tau \right)  \]

The right of the figure shows a blue $88 \times 88$ matrix $C$, with the $C_{14,15}$-th entry corresponding to the cross correlogram to the left. \\
 
Previous approaches to this type of problem rely on eigenvalues, which are invariant under change of basis and thus not an ideal solution as structure in neural data should be invariant under transformations of the form $C_{ij}=f(A_{ij})$ for a monotonically increasing function $f$.\\

Eigenvalue based approaches utilize symmetric eigenvalue decomposition to generate a set of eigenvalues that can be further analyzed. For example, eigenvalues were calculated for a random, symmetric $N \times N$ matrix were observed to follow the normal distribution. This sort of analysis is the foundation of the traditional eigenvalue approaches that this paper improves upon.\\

The limitations of eigenvalue based analysis is illustrated in the following figure. Note the destruction of spectral signatures for matrix structure by nonlinear monotonically increasing transformations. \\
\vfill

\pagebreak

\enlargethispage{2cm}
\begin{figure}[h!]
\centerline{\includegraphics[scale=0.45]{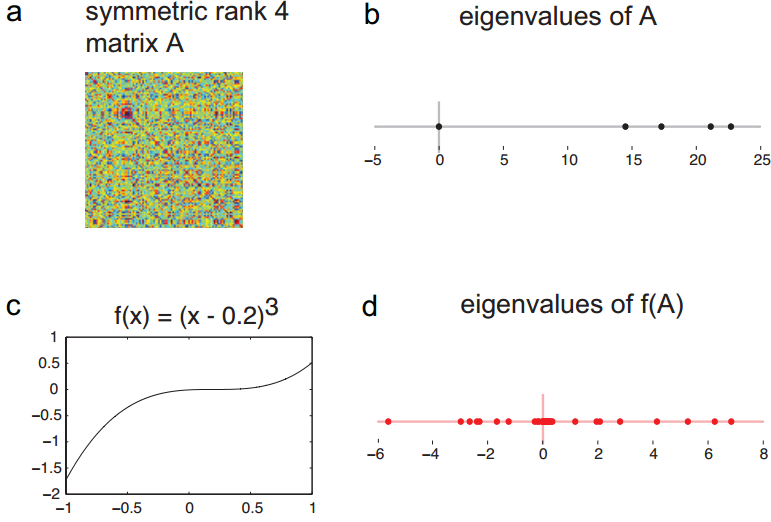}}
  \caption{Spectral signatures of matrix structure are destroyed by nonlinear monotonically increasing transformations.}
\end{figure}

(\textbf{a}) Consider a $100 \times 100$ symmetric matrix $A$ with $\texttt{rank}(A) = 4$. (\textbf{b}) The spectrum of $A$ includes four positive eigenvalues, indicating that $A$ has rank 4 and is positive semidefinite. The graph (\textbf{c}) depicts ${f(x) = (x-0.2)^3}$, a monotonically increasing function. The spectrum of the matrix $f(A)$ is shown in (\textbf{d}). Note that it contains many nonzero eigenvalues. The spatial signature that $A$ has a low-rank structure has been destroyed by $f$.

\section*{Methods and Analysis}
Observing structure is critical to understanding the organization and function of the underlying neural circuits -- structure may reflect the coding properties of neurons, rather than their physical locations. Geometric structure is expected here due to the existence of spatially localized receptive fields (place fields), but has not been previously detected \textit{intrinsically} using pattern correlations alone. A key question addressed by this work is whether the spatial coding properties of place cells is \textit{sufficient} to account for the observed geometric organization of correlations during spatial navigation. Alternatively, does this structure reflect finer features of the correlations, beyond what is expected from the place cells alone?\\

\vfill

\pagebreak

\enlargethispage{3cm}

To demonstrate successful extraction of geometric features using clique topology, data was taken from the pyramidal neurons of a rat hippocampus during a variety of activities including REM sleep, running, and spatial navigation. These neurons function as position sensors, firing at a high rate when the animal's position lies within the neuron's \textit{place field} -- its preferred region of the spatial environment. This sort of organization leads to a decrease in pairwise correlation $C_{ij}$ between place cells  as a function of the distances between the centers of place fields.\\ 

Place fields $F_i(\mathbf{x})$ were computed for each place cell with a two-dimensional spatial trajectory $\mathbf{x}(t)$. Synthetic spike trains were generated for each neuron as inhomogeneous Poisson processes with rate functions ${r_i(t) = F_i(\mathbf{x}(t))}$ given by the simple place field model. The synthetic spike trains preserved the influence of place fields while suppressing all other features, including non-spatial correlations. Betti curves derived from the place field model expressed all of the signatures for geometric organization. \\

The necessity of place field \textit{geometry} was also examined to determine if the Betti curves observed during spatial navigation could be attributed to the global signal that drives each neuron, $\mathbf{x}(t)$, filtered by a cell-specific function $F_i(\textbf{x})$. Place field data was scrambled by permuting the values of $F_i(\textbf{x})$ inside ``pixels'' of a $100 \times 100$ grid to produce non-geometric receptive fields $\widetilde{F}_i(\mathbf{x})$. Spike-trains were generated using the scrambled place fields $\widetilde{F}_i(\mathbf{x})$, and it was found that the second and third Betti curves lacked indication of geometric organization. This is illustrated in Figure 6.
\begin{figure}[h!]
\centerline{\includegraphics[scale=0.35]{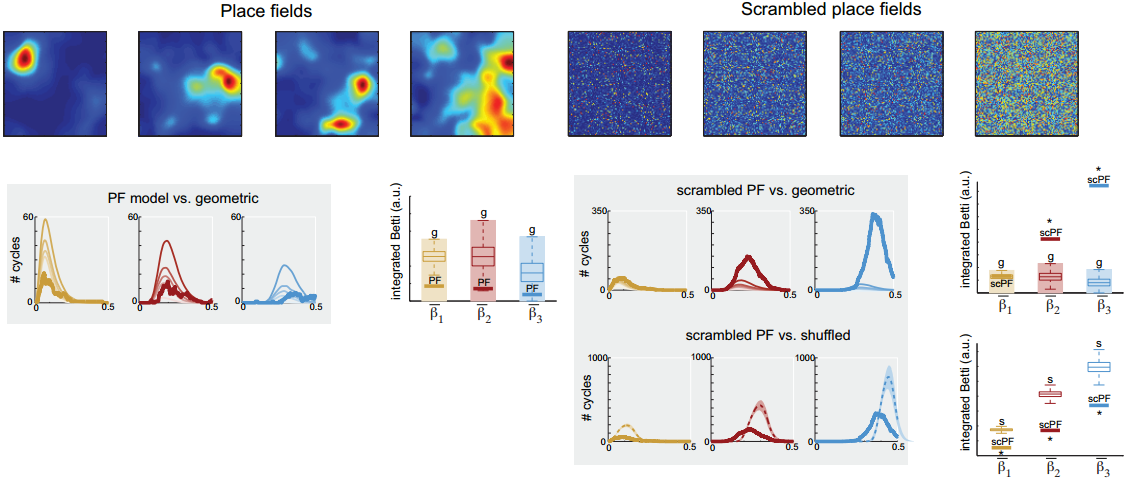}}
  \caption{Scrambled vs non-scrambled place fields with associated Betti curves and integrated Betti curves.}
\end{figure}

\pagebreak

Gusti et al. conclude that geometric signatures observed during spatial navigation reflect the geometry of place fields, and are not simply a consequence of neurons being driven by the global signal $\mathbf{x}(t)$. This suggests that geometric structure in place cell correlations is a consequence of positional coding and is not expected during non-spatial behaviors, however this notion was dismissed upon further analysis of non-spatial conditions (wheel running and REM sleep). It was found that Betti curves were again non-random and were consistent with signatures of geometric organization. \\

\begin{figure}[h!]
\centerline{\includegraphics[scale=0.3]{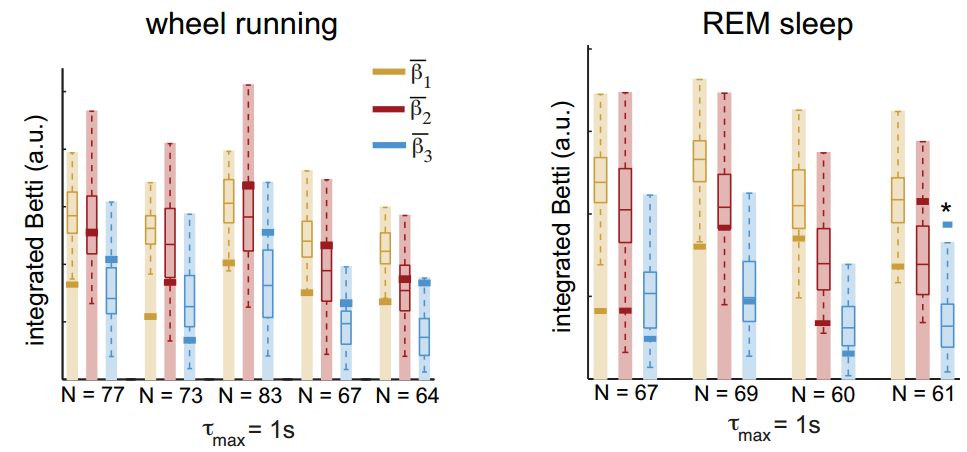}}
  \caption{Geometric organization in hippocampus during non-spatial behaviors.}
\end{figure}

The geometric organization in the non-spatial Betti curves is shown in the figure above. Please refer to the original publication for a more detailed explanation of the figure. 

\section*{Results}
Neural connections have many unknown nonlinearities, therefore clique topology must be used in lieu of eigenvalues to preserve nonlinear features present in the connectivity data. Using pairwise correlations of neurons in the hippocampus, Gusti et al. demonstrate that clique topology can be used to detect intrinsic structure in neural activity that is invariant under nonlinear monotone transformations without appealing to external stimuli or receptive fields. This demonstration serves to verify their newly discovered technique and represents a significant contribution towards the analysis of nonlinear systems. 

\section*{Critiques and Extension}
More general uses of clique topology are not discussed in the original publication. For example, Gusti mentions that the symmetric eigenvalue decomposition method is often used with success in physics applications, however there are many nonlinear systems in physics that could potentially benefit from analysis using clique topology. \\

The Van der Pol oscillator is a non-conservative oscillator that undergoes nonlinear damping, evolving according to the second order differential equation ${d^2x \over dt^2}-\mu(1-x^2){dx \over dt}+x= 0$. This equation has also been extended to model action potentials of neurons. It would be interesting to see clique topology applied to physical and chemical systems of nonlinear oscillators to detect some form of ``structure'' or pattern within the system.\\ 

\begin{figure}[h!]
\centerline{\includegraphics[scale=0.55]{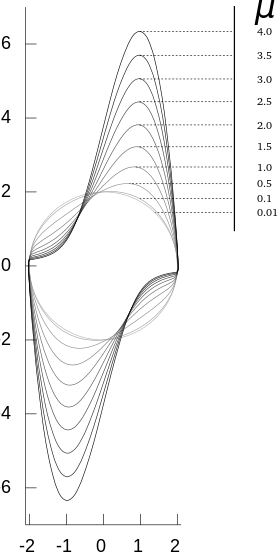}}
  \caption{Evolution of the Van der Pol oscillator limit cycle in the phase plane.}
\end{figure}

\section*{References}

\setlength{\parindent}{0cm}
Cartwright, M.L., ``Balthazar van der Pol", J. London Math. Soc., 35, 367-376, (1960).\\

C. Giusti, E. Pastalkova, C. Curto, and V. Itskov. Clique topology reveals intrinsic geometric structure in neural correlations. Proc. Natl. Acad. Sci., 112(44):13455–13460, 2015.\\

FitzHugh, R., Impulses and physiological states in theoretical models of nerve membranes'', Biophysics J, 1, 445-466, (1961).

\end{document}